# Computational engineering of sublattice ordering in a hexagonal AlHfScTiZr high entropy alloy

Lukasz Rogal[1a], Piotr Bobrowski[1], Fritz Körmann[2], Sergiy Divinski[3], Frank Stein[4], Blazej Grabowski[4b]

[1]*Institute of Metallurgy and Materials Science of the Polish Academy of Sciences, 30-059 Krakow, Poland*
[2]*Materials Science and Engineering, Delft University of Technology, 2628 CD Delft, Netherlands*
[3]*Institute of Materials Physics, University of Münster, Wilhelm-Klemm-Str. 10, 48149 Münster, Germany*
[4]*Max-Planck-Institut für Eisenforschung GmbH D-40237 Düsseldorf, Germany*

*Corresponding authors:* [a]*L. Rogal, l.rogal@imim.pl, tel.+48 122952801, fax +48 122952804*
[b]*Blazej Grabowski, b.grabowski@mpie.de, tel:+49 211 6792 512, fax:+492116792512*

**Abstract:** Multi-principle element alloys have enormous potential, but their exploration suffers from the tremendously large range of configurations. In the last decade such alloys have been designed with a focus on random solid solutions. Here we apply an experimentally verified, combined thermodynamic and first-principles design strategy to reverse the traditional approach and to generate a new type of hcp Al-Hf-Sc-Ti-Zr high entropy alloy with a hitherto unique structure. A phase diagram analysis narrows down the large compositional space to a well-defined set of candidates. First-principles calculations demonstrate the energetic preference of an ordered superstructure over the competing disordered solid solutions. The chief ingredient is the Al concentration, which can be tuned to achieve a $D0_{19}$ ordering of the hexagonal lattice. The computationally designed $D0_{19}$ superstructure is experimentally confirmed by transmission electron microscopy and X-ray studies. Our scheme enables the exploration of a new class of high entropy alloys.

Alloy design by combining multiple elements in near-equimolar ratios has the potential of creating new, unique engineering materials, commonly known as high entropy alloys (HEAs) or multi-principal element alloys [1-5]. HEAs have been shown to have excellent mechanical properties [6-12], as well as interesting magnetic [13-16] and electronic properties [17,18]. In contrast to conventional alloys, HEAs contain typically five or more elements [19-22] with concentrations ranging from 5 to 35 at.% [1]. The original idea behind this concept is to maximize the configurational entropy to achieve a single phase disordered solid solution. Such concentrated disordered solutions have been shown to exist on the face-centered cubic (fcc) and body-centered cubic (bcc) lattice [1-3], and more recently also on the hexagonal close-packed (hcp) lattice [23-25]. Theoretical approaches based on first principles were proposed [26,27] to replace the earlier empirical rules and to guide the search for disordered solid solutions.

Here we propose a *reverse* design strategy, with the aim of finding *ordered* structures in HEAs. This is motivated by the fact that—in contrast to fully disordered alloys—the ordered counterparts usually work-harden faster showing an improvement in their mechanical properties [28,29]. The increased work-hardening rate of ordered alloys originates from a high storage rate of dislocations coupled with a general lack of dynamic recovery processes [29]. A prominent example of an outstanding impact of ordered phases leading to ultra-high strength are $Ti_3Al$-base alloys [30]. In HEAs, ordered phases have been reported in bcc structures containing elements with a large electronegativity difference such as Al and Ni [31-34]. For example, increasing the Al concentration of fcc/bcc based Al$_x$CoCrCuFeNi alloys leads to chemical ordering, which correlates with an increase in hardness [1], although ductility and fracture toughness suffer typically [35]. Ordered hcp based phases in HEAs have not been found so far. Such a new class of ordered hcp HEAs could potentially open new routes to improving materials properties, as may be anticipated from the beneficial properties of the hcp based "Super-Alpha-2" Ti$_3$Al-base alloys [30].

A note is due here. With an ordered HEA we do not mean that all elements occupy own sublattices as is typically the case in binary alloys. In HEAs, several elements can share the same sublattice with a preference of some of the elements to a specific sublattice, e.g., A and D atoms may occupy, say, the α-sublattice and C, D, and E atoms reside on the β-sublattice in an ordered ABCDE HEA. Or, just one atomic species could reveal a preference to a given sublattice whereas all other atoms may be equally distributed. As this comment should make clear, a large number of potential types of ordering are possible in HEAs, thereby opening new perspectives for alloy optimization and development.

The fundamental challenge in developing multi-principle element alloys is the immense combinatorial amount of possible phases with varying composition. To overcome this challenge, sophisticated approaches and design principles are required [36,37]. In this work we demonstrate how a combination of thermodynamic arguments, first principles calculations, and experiments can be used to narrow down the *a priori* large range of potential candidate structures and to find a material with the desired property. We apply this strategy to the Al-Hf-Sc-Ti-Zr system with the goal of finding a new HEA with an ordered D0$_{19}$ superstructure on the hcp lattice.

**Results**
**Thermodynamic analysis.** In an ideal case, to find the desired ordered hcp superstructure for our Al-Hf-Sc-Ti-Zr HEA, we would perform a quantitative, CALPHAD [38,39] based stability analysis of the relevant phases. Unfortunately, CALPHAD parametrizations of multi-principal element alloys, as the considered five-component HEA, are unavailable as yet. This is nowadays a typical restriction for the design of HEAs. Moreover, even if a parametrization was available, it could not be guaranteed that it provides a reasonable description of the HEA system for which it has (in the usual case) not been optimized. We therefore employ the CALPHAD approach only in a qualitative manner, which nevertheless turns out to be a very helpful strategy to narrow down the candidates for the *ab initio* simulations. We employ in particular a strategically useful representation and analysis of the sub-binary phase diagrams of our system. The advantage is that all of these phase diagrams are determined experimentally and can be straightforwardly obtained from databases.

As shown in Fig. 1, except for the Al-containing binaries (top row) the other phase diagrams reveal a large solubility in the disordered bcc A2 phase (red regions) at higher temperatures, and mostly also in the disordered hcp A3 phase (light green) at lower temperatures. The good miscibility can be understood by considering the similar nature of the involved *d* transition elements, like the atomic volume or the electronegativity (Tab. 1). A closer look at Tab. 1 reveals, however, that Ti is somewhat special, having a smaller atomic volume and a larger electronegativity than Sc, Hf, and Zr. The special character of Ti is in fact also reflected by the binary phase diagrams of the $d$ elements (Fig. 1 without the top row), which show, e.g., that the Sc-Ti system features a miscibility gap in the hcp A3 phase, decomposing into a hcp A3+hcp A3 two-phase field. The Hf-Ti system also shows a tendency to an hcp A3+hcp A3 decomposition at lower temperatures, and in fact this applies also to the Ti-Zr system below the shown 400 K.

Adding Al to the $d$ transition elements changes the appearance of the binary phase diagrams completely (top row of Fig. 1). Al is a strong intermetallic former as is clearly reflected by its phase diagrams showing various intermetallic phases (grey regions). All of the Al phase diagrams show for example an $Al_3M$ phase (M=Hf, Sc, Ti, Zr) which forms a two-phase field with the fcc A1 phase towards the Al-rich side of the phase diagrams (grey-orange stripes). Almost all of the formed intermetallic phases are stoichiometric phases, i.e., they exist only at a single, fixed composition and have no solubility. The Al-Ti diagram is an exception. The $L1_0$ phase which forms at around Al50:Ti50 on an fcc lattice has an appreciable solubility range. The other, very special and for the present analysis important phase is the ordered $D0_{19}$ phase (dark green region) which forms at around Al25:Ti75 on the hcp lattice. It has a similarly large solubility range as the $L1_0$ phase, extending towards both, the Al-rich and Ti-rich side. The $D0_{19}$ phase is embedded into an hcp A3 solid solution (light green region), which extends even to a composition of 50:50 at higher temperatures. This solubility range is much larger than for any of the other shown Al phase diagrams, again indicating the special character of Ti among the investigated $d$ elements. Tab. 1 shows that the atomic volume and electronegativity of Ti are indeed closer to the ones of Al than to the ones of the other $d$ elements.

In order to elucidate (at least qualitatively) how the binary information can be extended to the five-component Al-Hf-Sc-Ti-Zr HEA, we combined the binary diagrams into a 3D-representation as shown in Fig. 2. Joining all diagrams from Fig. 1 is not feasible and we therefore restricted the diagram to a quaternary sub-diagram, keeping the most relevant elements for our purpose, Al (at the top) and Ti (to the left). An important feature that we have added to the 3D-phase diagram is the $AlM_3$ plane where M=Ti, Sc, Hf, Zr. Moving on this plane towards the Ti-Al phase diagram we approach the $D0_{19}$ phase. Given this representation and the fact that—despite the discussed differences—Ti does feature chemical similarity with the other $d$ elements, one can speculate that the ordered $D0_{19}$ phase extends into the multi-component system as indicated by the dark green arrow in Fig. 2.

In Ref. [40] the equiatomic HfScTiZr multi-principle element alloy was studied and it was shown that it consists of a disordered hcp A3 solid solution. Taking this alloy as a reference (black dot labeled "0 Al" in Fig. 2) and introducing Al into this system with a concentration of >5 at.%, allows us to stay within the usual definition of HEAs. The only design criterion left is which element(s) to remove for Al. From the above analysis we know that Ti is a critical factor for the formation of the $D0_{19}$ phase and we thus keep its concentration. Out of the remaining $d$ elements Sc shows the largest tendency to phase decomposition (Sc-Ti phase diagram), which is related to the large atomic radius and electronegativity of Sc (Tab. 1). This special feature of Sc was responsible for the presence of Sc enriched phases and a strong segregation of Ti to interfaces in HfScTiZr [40]. We therefore substitute in the subsequent computational modeling Sc by Al, in particular taking concentration steps of 5 at.% Al (black points "5 Al", "10 Al", and "15 Al" in Fig. 2), approaching thereby the AlM$_3$ plane and thus the possibility of D0$_{19}$ ordering.

***Ab initio* calculations of the phase stabilities.** In order to quantitatively assess the phase stabilities for the selected compositions, we resort in the following to unbiased and very accurate, parameter-free first-principles calculations realized by DFT. In particular we investigate whether an ordered D0$_{19}$ superstructure can be energetically stable over the fully disordered hcp A3 and bcc A2 solid solutions, which are the most relevant phases according to the above phase diagram analysis. Phase decomposition could be in principle also investigated but is very difficult in practice due to the large configuration space. Therefore, our computations were restricted to energy differences at fixed compositions. In this respect, we note that D0$_{19}$ and A3 phase are located on the same hcp lattice, the difference between them being related to the chemical order/disorder.

To resemble this situation as closely as possible we have generated large special quasi random structures (SQS). Fig. 3a and b show the SQS supercell representing the hcp A3 phase with full chemical disorder. Fig. 3c and d show the supercell representing the D0$_{19}$ phase. The D0$_{19}$ like supercell requires a close look to understand where the ordering occurs. Note for that purpose the small red balls attached to some of the atoms in Fig. 3c and d, which indicate the sublattice positions of the Al atoms if the composition was strictly AlM$_3$ (M=Ti, Sc, Hf, Zr). For our selected compositions, there are not enough Al atoms to fill up all these sublattice sites, and thus this sublattice is only *partially* ordered. The remaining sites of the Al sublattice must be filled up by the other constituents and we have assumed that all other atoms can enter the Al sublattice and that there is no ordering among them. The other sublattice contains the *d* elements only and is fully disordered. To guarantee convergence with respect to chemical disorder we tested two distinct SQS supercells for each of the phases, and to match the experimental boundary conditions we optimized the volume, *c/a* ratio, and local relaxations for D0$_{19}$, hcp A3, and bcc A2 separately.

The resulting *ab initio* energetics for the example of Al$_{15}$Hf$_{25}$Sc$_{10}$Ti$_{25}$Zr$_{25}$ at.% is summarized in Fig. 4. Fig. 4a shows the energy contour plot for the D0$_{19}$ phase as a function of atomic volume and *c/a* ratio. The two main conclusions are: (1) The energy dependence on the *c/a* ratio is small. In the investigated *c/a* range (1.57…1.63) the energy at the equilibrium volume changes only by 2 meV/atom. (2) The equilibrium *c/a* ratio and the equilibrium volume are independent of each other. For example, for all volumes the equilibrium *c/a* ratio is 1.60 as indicated by the black dashed line. Both conclusions apply likewise to the hcp A3 phase, except for the fact that the equilibrium *c/a* ratio is slightly smaller with 1.59 (red dashed line). Tab. 2 shows that the computed values agree well with our experimental results. Fig. 4b compares the $T=0$ K energies of all investigated phases revealing clearly that the D0$_{19}$ phase is the most stable one. The variation in the energy difference due to the different SQS supercells for D0$_{19}$ and hcp A3 is reasonably small and we can conclude that the partially ordered D0$_{19}$ superstructure is about 30 meV/atom more stable than the fully disordered hcp A3 structure and about 60 meV/atom more stable than the disordered bcc A2 at $T=0$ K.

In order to determine the temperature dependent phase stabilities, we computed finite temperature contributions due to electrons, atomic vibrations, and configurational entropy.

The electronic and vibrational Gibbs energies are similar for all three phases and contribute therefore little to the differences as exemplified by the small light gray and light green shaded areas in Fig. 4c for $D0_{19}$ vs. A3. A similar observation was recently made for other HEAs [41]. The dominant contribution to the Gibbs energy difference comes from the configurational entropy (blue shaded area). The reason is that the configurational entropy is considerably smaller for the $D0_{19}$ phase, because the Al atoms are confined to the Al sublattice in contrast to the hcp A3 phase where the Al atoms can be located on any of the hcp lattice sites. Therefore, the stability of the disordered hcp A3 phase is more and more increased as temperature rises, leading eventually to a phase transition from $D0_{19}$ to A3. Averaging over both SQS supercells for each phase gives a transition temperature of about 1230 K. Note that by assuming an ideal entropy [Eqs. (5-11)] and neglecting short-range order, the predicted order-disorder temperature is likely underestimated [41,42]. Considering the variation in the energies of the SQS supercells (Fig. 4b) provides in fact a temperature window of 1100…1350 K for a possible transition as indicated by the gray diagonal stripes in Fig. 4c. The stability of the bcc A2 is likewise increased as compared to $D0_{19}$, but—due to the considerable $T = 0$ K offset—bcc A2 cannot compete with the other phases at any of the investigated temperatures.

Reducing the Al concentration (black dots in Fig. 2) causes a decrease of the order-disorder transition temperature, as already anticipated based on the phase diagram analysis. This decrease is, however, not very strong. For $Al_5Hf_{25}Sc_{20}Ti_{25}Zr_{25}$ at.% the transition temperature is reduced only by about 200 K to 1000 K. The reason for the rather weak reduction of the order-disorder transition temperature is found in the detailed balance between the $T=0$ K energy and the thermal entropy difference. The $T=0$ K energy difference between the partially ordered $D0_{19}$ superstructure and the disordered hcp A3 phase is indeed considerably smaller for the $Al_5Hf_{25}Sc_{20}Ti_{25}Zr_{25}$ at.% composition ($Al_5$: +6 meV/atom vs. $Al_{15}$: +30 meV/atom). But, the difference in the configurational entropy likewise decreases because the $Al_5Hf_{25}Sc_{20}Ti_{25}Zr_{25}$ alloy has even less order and thus more configurational disorder than the $Al_{15}Hf_{25}Sc_{10}Ti_{25}Zr_{25}$ one.

**Microstructure characterization.** Based on the preceding theoretical analysis we have selected the composition of $Al_{15}Hf_{25}Sc_{10}Ti_{25}Zr_{25}$ at.% for a detailed experimental investigation. The SEM image of the corresponding as cast sample (Fig. 5a) indicates a two-phase microstructure with a fine plate arrangement, consisting of laths with a bimodal size distribution: coarse laths with 2-30 μm in length and 0.3-0.8 μm in thickness, and fine laths of 0.5-4 μm in length and 0.01-0.3 μm in thickness, and areas of a second phase between the laths (the interspace region with dark contrast). This microstructure morphology is similar to the α' martensitic one in rapidly cooled Ti-6Al-4V wt.% alloys [30].

The TEM bright field image in Fig. 5b shows contrast from a few large laths (bright regions) with a thickness of 0.3-0.5 μm into which smaller laths of 20-100 nm in size are embedded. The concentrations shown in Tab. 3 (obtained from TEM-EDS) indicate a rather homogenous distribution of the involved elements in the matrix with small segregation of Al to the interspace region. This small concentration variation suggests the presence of two types of phases, consistent with the SEM results (Fig. 5a).

The X-ray analysis of the as cast alloy (lower panel in Fig. 6) indicates—except for a small amount of a primary (Sc,Zr)$_2$Al phase with hexagonal structure—the presence of a hexagonal solid solution with lattice constants of $a$ = 3.1092 Å and $c$ = 4.9271 Å (see also Tab. 2) and space group *P6$_3$/mmc*. The SAED pattern shown in the inset of Fig. 5b is consistent with the X-ray results, showing strong reflections at the standard hcp positions of (2021) and (02-22). However, a closer inspection of the SAED pattern also shows an additional weak reflection at (01-11) and (21-30) (highlighted by the circles in the inset of Fig. 5b), which is an indication of the onset of *ordering* on the hcp lattice. Since the X-ray analysis does not indicate any ordering, we conclude that the partial ordering in the as cast state is weak and can be only detected by the more sensitive SAED method.

In order to determine the effect of temperature on the microstructure of the Al$_{15}$Hf$_{25}$Sc$_{10}$Ti$_{25}$Zr$_{25}$ HEA, a DTA analysis was conducted (Fig. 7). The first heating curve of the as-cast sample (black line) reveals several small exothermal peaks below 800 °C (cf. zoom in the inset) related to the transition from the meta-stable as-cast state towards the equilibrium state. A larger endothermal peak is present at around 950°C presumably caused by dissolution of the (Sc,Zr)$_2$Al precipitates. The two weak, exothermal peaks below 800 °C completely vanished after the first heating and cooling cycle. Similarly, the peak around 950 °C continuously became smaller and vanished after some heating/cooling cycles indicating that the (Sc,Zr)$_2$Al precipitates were meta-stable. The red line in Fig. 7 shows the heating curve after six times heating and cooling, where no more changes occurred in the curves indicating that the material was in an equilibrated state. A new endothermal peak is visible near 900 °C. We identify this peak as corresponding to the order-disorder transition from the partially ordered D0$_{19}$ superstructure to the disordered hcp A3 phase predicted by our *ab initio* calculations. The large peak at around 1500°C corresponds to the melting transition [solidus=1442(±2)°C and liquidus=1500(±5)°C]. Since neither our calculations have predicted the disordered bcc A2 phase to be stable nor does our DTA curve show another peak before melting, we conclude that our HEA melts directly from the hcp A3 phase. Based on the knowledge of the DTA results we conducted an annealing of the as-cast sample at 1000°C for 5h, leading to plate coarsening of the solid solution (average size of the laths reached tens of microns), as well as to further homogenization of the chemical composition.

The EBSD analysis of the lattice misorientations in the annealed sample revealed that a large fraction of grain boundaries is characterized by 60° and 90° misorientations, which constituted, respectively, 61% and 7% of the total boundary length (marked as black and white lines, respectively, in Fig. 8a). Additionally, Σ19 boundaries were also identified in the analyzed material. These results indicate that during annealing new grains, which maintain crystallographic orientation relationships, are nucleating at the edges of the existing grains. Subsequently, the microstructure coarsening proceeds through grain growth. In addition to the coarsening, we further identified small Sc, Zr enriched secondary precipitation particles with an irregular shape and 0.1-1.5 μm in size based on the EBSD map (small dark area, in Fig. 8a) and the electron diffraction results.

The TEM-BF image of the annealed sample (Fig. 8b) shows a large lath with strong D0$_{19}$ superstructure reflections (SAED pattern in the inset) and small precipitates of the Sc enriched phase. A volume ordering effect in the annealed sample was confirmed by X-ray analysis, where additional peaks in the range of low 2θ values were identified (marked with a circle in Fig. 6). Significant differences between peak intensities were observed (100, 002, 101 – A3 and 200, 002, 201 – $D0_{19}$). This suggests that the annealing led to a rearrangement of the atoms in the lattice. Additionally, small shifts of characteristic peaks to lower 2θ values were observed due to homogenization of the structure and dissolution of the $(Sc,Zr)_2Al$ precipitates (identified before in the as cast alloy and confirmed by the DTA analysis), which partially expanded the lattice.

**Mechanical properties.** The average hardness of the $Al_{15}Hf_{25}Sc_{10}Ti_{25}Zr_{25}$ HEA in the as cast state was 524 HV and it decreased after annealing to 407 HV. This decrease is probably related to the coarsening of the microstructure and the strain release in the annealed sample. Compressive stress-strain plots are shown in Fig. 9a for the as-cast state (curve 1) and after annealing at 1000°C/5h (curve 2). In the as-cast state, the sample shows a high yield strength of 1450 MPa and a high compression strength of 1950 MPa at a strain of 4.6%, fracturing in a brittle manner. Annealing decreases the yield strength down to 980 MPa and increases the compressive strength, even up to 2200 MPa at an appreciable strain level of 21.5 %.

The bright field TEM micrograph of the alloy after compressive deformation to about 22 % plastic strain exhibits multiple shear bands of 50 to 100 nm thickness, with a high defect density (Fig. 9b). The SAED pattern, taken from a shear band region, shows reflections of the hexagonal superstructure with [1-100] zone axis. These observations confirm that shear band formation represents the key mechanism of deformation in the $Al_{15}Hf_{25}Sc_{10}Ti_{25}Zr_{25}$ HEA. In contrast, the dominant deformation mechanism of pure hcp based Mg, Ti, and Zr materials involves twinning rather than slip. It is known that slip leads to high local stress concentrations, causing an increase in hardness [43-45]. We believe that the shear band deformation is responsible for the substantial compressive plasticity of our ordered HEA with hexagonal superstructure. Additionally, the compressive strength of the chemically disordered $Hf_{25}Sc_{25}Ti_{25}Zr_{25}$ HEA was measured for comparison. It can be seen that adding aluminum to the Hf-Sc-Ti-Zr system improves both, compression and yield strength.

In *typical* alloys an ordered structure leads to an *increase* in hardness and a *decrease* in ductility, see e.g. [35]. For example Ti-Al alloys with non-symmetrical (tetragonal) $DO_{22}$ crystal structure and with the $D0_{19}$-type $Ti_3Al$ phase are hard and brittle due to an insufficient number of independent slip systems for plastic deformation [46]. The situation is *reversed* in the present partially ordered hexagonal HEA, where compressive plasticity increases significantly with slight increases in yield strength. A higher strength and higher ductility were also observed in a Ti-Al system alloyed with Nb [47]. One of the possible reasons for the unusual compressive properties of the ordered structure in the present HEA could be the increased number of elements in the system, leading to a plasticity and strength increase due to the reduced slip planarity and an increased non-basal slip activity, unlike in $Ti_3Al$ [47]. Our findings underline that conclusions on ordering in alloys derived from mechanical properties, in particular for HEAs, need to be carefully validated.

In order to evaluate to what extent these conclusions can be transferred to tensile deformation, we performed micro-mechanical tensile tests using a custom-built device [48]. Dog-bone shaped samples with a gauge length of about 3 mm and 1.0×0.3 $mm^2$ cross section were cut by spark erosion from the as-cast and annealed $Al_{15}Hf_{25}Sc_{10}Ti_{25}Zr_{25}$ HEAs. The results of the tensile tests are shown in the supplementary material. Whereas a tensile elongation of about 0.5% was observed for the as-cast sample, the annealed sample broke in the elastic region at a stress of about 1000 MPa. Correspondingly, fundamentally different fracture surfaces were observed in these two states, namely a predominantly ductile fracture with a combination of transgranular and intergranular cracks in the as-cast state and an explicit transgranular brittle fracture in the annealed sample.

Nevertheless we can state that already 0.5% of tensile ductility in the as-cast state is a highlight for an hcp-based HEA in view of the practical absence of similar studies for, e.g., titanium aluminides. The partially ordered $Al_{15}Hf_{25}Sc_{10}Ti_{25}Zr_{25}$ HEA with distinct sublattices does not reveal an enhanced tensile ductility and broke in brittle manner. However, the fracture surface reveals traces of significant dislocation activity.

These findings support our idea of a potential application of ordered HEAs as, e.g., strengthening particles in a hcp matrix.

**Discussion**

The original design strategy of multi-component high entropy alloys forming disordered random solid solutions is reversed. In contrast to the *traditional* scheme we specifically design *ordered* or *partially ordered* HEA superstructures. This is achieved by combining thermodynamic CALPHAD and *ab initio* calculations. A CALPHAD approach is employed to narrow down the large configuration space to a distinct series of Al-Hf-Sc-Ti-Zr alloys. Subsequent quantitative *ab initio* calculations demonstrate that the presence of aluminum in the $Al_{15}Hf_{25}Sc_{10}Ti_{25}Zr_{25}$ at.% high entropy alloy can lead to the formation of a partially ordered $D0_{19}$-type hexagonal superstructure. The computationally predicted target compositions are experimentally verified. The microstructure of our alloy showed thermal stability during annealing at 1000°C/5h. In contrast to typical ordered alloys our compression strength studies revealed that the ordered hexagonal superstructure has high ductility. In tension, we found a lack of ductility. We highlight nevertheless that titanium aluminides are usually characterized by an absence or significantly reduced tensile ductility and the value of 0.5% tensile elongation to fracture in the present as-cast HEA is remarkable. We expect that the full potential of the ordered HEA phase could be exploited by optimizing its microstructure.

A strong impact of aluminum on high entropy alloys with hexagonal structure, with the possibility of superstructure formation, has been demonstrated. Our approach is general and can be used to design new HEAs with other types of ordered superstructures. Moreover we open the route to the new and hitherto unexplored class of ordered hcp HEAs, of which binary counterparts are already well established and highly successfully applied in industrial applications.

**Methods**

**First principles.** We employed first-principles calculations within density-functional theory (DFT) as implemented in the VASP software [49,50] in conjunction with the projector-augmented wave method [51] and the generalized gradient approximation in the Perdew-Burke-Ernzerhof parametrization [52]. A plane wave cutoff of 240 eV was chosen for all calculations. Chemical disorder was simulated by special quasi random structures (SQS) [53]. For both of the studied hcp phases, D0$_{19}$ and A3, 216 atom supercells were employed, corresponding to 6x6x3 times the primitive hcp cell. For both phases two distinct SQS structures were tested to assess the performance in representing the chemical disorder. Dense *k*-point grids of up to 5x5x5 were studied (27,000 *k*-point•atom) to ensure high numerical precision. The volume and *c/a* ratio of the supercells were optimized by computing total energies for a grid of 4 different *c/a* ratios (1.57, 1.59, 1.61, 1.63) for each of the 9 considered volumes chosen around the equilibrium volume. The 216 atom supercells contained 32 Al, 54 Hf, 22 Sc, 54 Ti, and 54 Zr atoms which well resembles the experimentally investigated Al$_{15}$Hf$_{25}$Sc$_{10}$Ti$_{25}$Zr$_{25}$ alloy. The Al$_5$Hf$_{25}$Sc$_{20}$Ti$_{25}$Zr$_{25}$ HEA was also represented with a 216 atom SQS supercell, but containing the following numbers of atoms: 11 Al, 54 Hf, 43 Sc, 54 Ti, and 54 Zr. The bcc A2 phase was investigated only for the Al$_{15}$Hf$_{25}$Sc$_{10}$Ti$_{25}$Zr$_{25}$ HEA. We used a 54 atom supercell corresponding to 3x3x3 times the conventional cubic unit cell. The *k*-point mesh was set to 6x6x6. The volume was sampled on 8 points around the equilibrium volume. Since in the chosen 54 atom bcc supercell the exact composition of the hcp supercells cannot be achieved, we used the following trick. We created two bcc SQS supercells with slightly varying compositions, such that the average over them gave the exact hcp composition (for the Al$_{15}$Hf$_{25}$Sc$_{10}$Ti$_{25}$Zr$_{25}$ HEA). In particular, one bcc SQS contained 8 Al, 13 Hf, 6 Sc, 13 Ti, and 14 Zr atoms, whereas the other one contained 8 Al, 14 Hf, 5 Sc, 14 Ti, and 13 Zr atoms. All *T* = 0 K calculations were performed with the Methfessel-Paxton technique [54] with a smearing value of 0.1 eV. The energy volume dependence was fitted to a Vinet equation of state [55].

In order to capture the effect of electronic excitations we employed the finite temperature extension to DFT by Mermin [56]. In particular, we computed the electronic free energy by [57,58]

$$F_{\text{el}} = U_{\text{el}} - TS_{\text{el}}, \qquad (1)$$

$$U_{\text{el}} = \int_{-\infty}^{\infty} D(\varepsilon) f(\varepsilon, T)\, \varepsilon\, d\varepsilon - \int_{-\infty}^{\varepsilon_F} D(\varepsilon)\, \varepsilon\, d\varepsilon, \qquad (2)$$

$$S_{\text{el}} = 2k_B \int_{-\infty}^{\infty} D(\varepsilon) s(\varepsilon, T)\, d\varepsilon, \qquad (3)$$

$$s(\varepsilon, T) = -[f \ln f + (1-f)\ln(1-f)], \qquad (4)$$

where $T$ is the temperature, $D(\varepsilon)$ the electronic density of states, $f = f(\varepsilon, T)$ the Fermi-Dirac function, $\varepsilon_F$ the Fermi energy, and $k_B$ the Boltzmann constant. Atomic vibrations were treated within the Debye formalism as described in detail in Ref. [59]. For the configurational entropy we assumed ideal mixing. In this approximation, the configurational entropy of the fully disordered bcc A2 and hcp A3 phase is given by

$$S_{\text{conf}}^{\text{bcc/hcp}} = -k_B \sum_i x_i \log x_i, \qquad (5)$$

where the $x_i$ are the element concentrations in the homogenously disordered alloy for the 5 different species *i*. For the partially ordered D0$_{19}$ phase, the configurational entropy can be decomposed into contributions for the two inequivalent sites A and B as

$$S_{\text{conf}}^{\text{D0}_{19}} = \tfrac{1}{4} S_{\text{conf}}^{\text{D0}_{19}-\text{A}} + \tfrac{3}{4} S_{\text{conf}}^{\text{D0}_{19}-\text{B}}, \qquad (6)$$

with

$$S_{\text{conf}}^{\text{D0}_{19}-\text{A,B}} = -k_B \sum_i y_i^{A,B} \log y_i^{A,B}, \qquad (7)$$

where the partial Al ordering results in two inequivalent sublattice concentrations of

$$y_i^{A,B} = w_i^{A,B} x_i, \qquad (8)$$

determined by weighting factors $w_i^{A,B}$. The sublattice concentrations render the multiplicities of the sublattice of 1 and 3 as

$$x_i = \frac{1}{4} y_i^A + \frac{3}{4} y_i^B, \tag{9}$$

fulfilling the sum rule, i.e., $\sum_i y_i^{A,B} = 1$. If Al is confined to the A-sublattice, $w_{Al}^A = 4$ and $w_{Al}^B = 0$, and assuming a homogenous distribution of the other elements (Hf, Sc, Ti, Zr), the weighting factors for the other four components are given by

$$w_{i \neq Al}^A = (1 - 4x_{Al})/(1 - x_{Al}), \tag{10}$$

and

$$w_{i \neq Al}^B = 1/(1 - x_{Al}). \tag{11}$$

Eqs. (5-11) provide the ideal configurational entropy of bcc A2, hcp A3, and D0$_{19}$ for any given composition $x_i$. The assumption of ideal configurational entropy (of mixing) results by definition in 1st order phase transitions and typically provides theoretical lower bounds of order-disorder transition temperatures [41,42].

**Experimental.** Our experimental analysis focused mainly on the composition of Al$_{15}$Hf$_{25}$Sc$_{10}$Ti$_{25}$Zr$_{25}$ at.% which shows the superstructure. An additional alloy with a composition of Al$_5$Hf$_{25}$Sc$_{20}$Ti$_{25}$Zr$_{25}$ at.% was prepared and analyzed to confirm that it contains no superstructure. The alloys were prepared from elements of 99.99 wt.% purity in an arc melting furnace with a water-cooled copper plate under a protective Ar atmosphere. For structural characterization, cross sections of drops were analyzed. The drops solidified with an average cooling rate of 300°C/s. X-ray measurements of the phase composition were performed using a Philips PW 1410 diffractometer and CoKα filtered radiation. Scans were performed on polished cross sectioned samples at 15° to 100°, with a step size of 0.02°, and a dwell time of 10 s. A Rietveld refinement was performed in order to determine the type of structure and lattice parameters. The microstructure was examined using a scanning electron microscope (SEM), FEI Quanta equipped with an energy-dispersive X-ray spectrometer EDAX Apollo and a EBSD camera EDAX Hikari. Samples for SEM studies were electropolished in an A2 reagent using Struers Lectropol-5. Further microstructure analysis including selected area electron diffraction (SAED) patterns was performed using the Tecnai G2 F20 transmission electron microscope (TEM). The micro-chemical analysis was conducted using the TEM in scanning transmission electron microscopy (STEM) mode coupled with Integrated Energy-Dispersive X-ray spectroscopy (EDS). Differential thermal analysis (DTA) was carried out with a Setaram SETSYS-18 DTA. A cylindrical sample of 3 mm in diameter and height was placed in an alumina crucible and measured under Ar atmosphere. The sample was heated and cooled seven times using the following heating/cooling rates: 10, 10, 20, 5, 10 °C/min up to 1200 or 1250°C and back down to room temperature, then 20 °C/min to 1250°C and continuing with 5 °C/min to 1580°C, the same back to room temperature and finally 10 °C/min to 1580°C and back. Calibration measurements were performed using certified standards of pure Al, Au, and Ni resulting in an accuracy of ±1 °C for the measured temperatures. Hardness measurements (with the Vickers method) were carried out using a Zwick/ZHU 250 (HV5) in accordance with (HV) ASTM E 92. Compressive strength tests were performed using an INSTRON 6025 machine according to the PN-57/H-04320 standard on cylindrical samples of 4 mm in diameter and 6 mm in height.

**Acknowledgments.** We thank Andrei V. Ruban for providing the spcm program for generating the special quasi-random structures. We thank M. Wegner and D. Gaertner for performing the tensile tests and fracture surface characterization. The research was supported by the Polish science financial resources, The National Science Centre, Poland, project title: "Development of new high entropy alloys with dominant content of hexagonal solid solutions" project number: 2014/15/D/ST8/02638. Funding by the European Research Council (ERC) under the European Union's Horizon 2020 research and innovation programme (grant agreement No 639211) and by the scholarship KO 5080/1-1 of the Deutsche Forschungsgemeinschaft (DFG) are gratefully acknowledged.

**Author contributions.** F.K. and B.G. performed the thermodynamic analysis and the *ab initio* simulations. L.R. prepared the alloys. L.R. performed the TEM and X-ray investigations. L.R. and S.D. performed the mechanical studies. P.B. performed the SEM and EBSD measurements. F.S. performed the calorimetric measurements. L.R., F.K., S.D., and B.G. prepared the manuscript. All authors approved the final manuscript for submission.

**Additional Information**
Competing financial interests: The authors declare no competing financial interests.

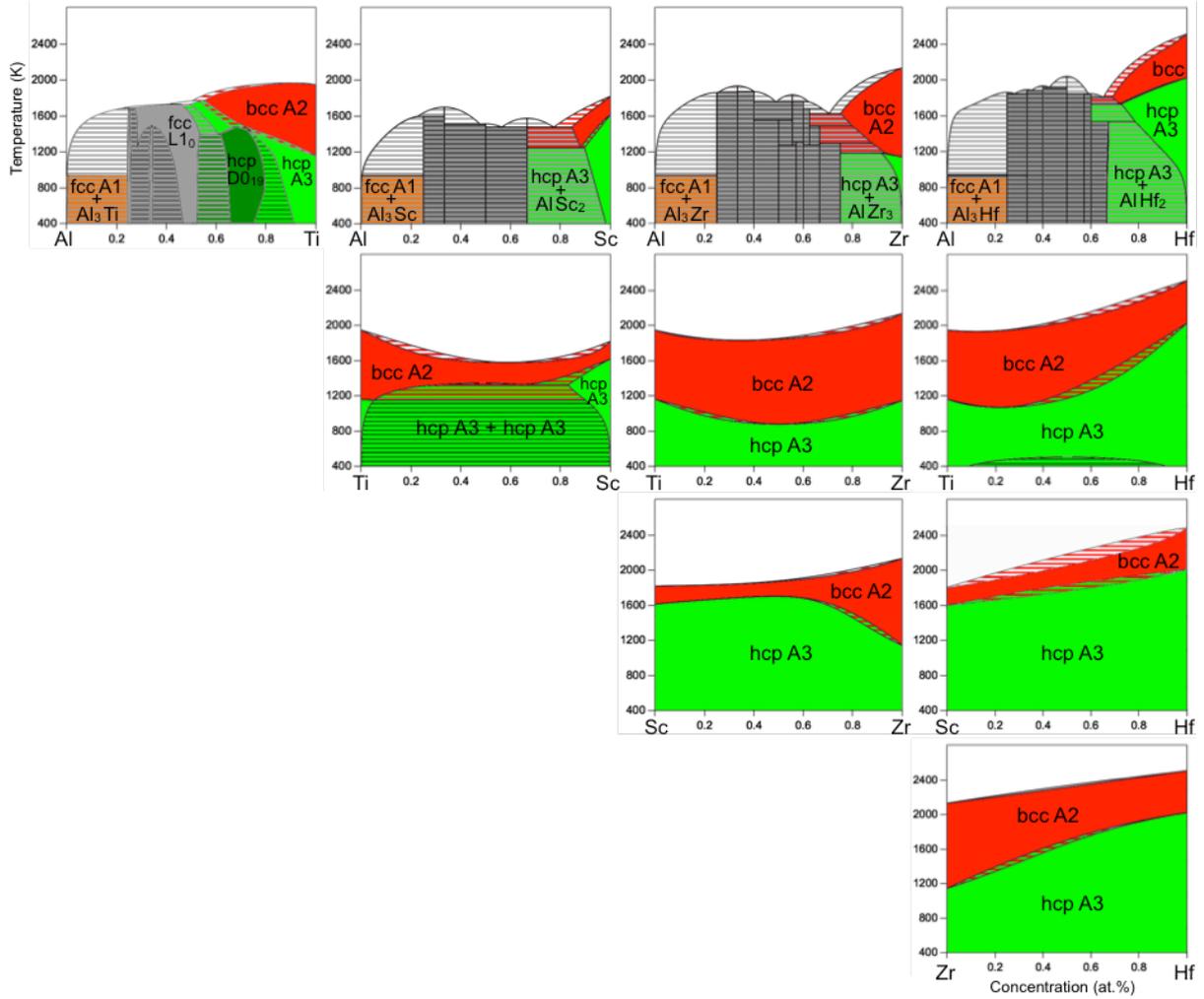

Fig. 1: Phase diagrams of all 10 individual constituting binaries of the considered Al-Hf-Sc-Ti-Zr HEA. The phase diagrams for Hf-Ti, Hf-Zr, and Ti-Zr are based on the SGTE (2014) alloy database. The phase diagrams for Al-Hf, Al-Sc, Al-Zr, Sc-Ti, and Sc-Zr are based on the PanSolution database (ipandat.computherm.com). The Al-Ti phase diagram corresponds to the reassessment of Schuster and Palm [60]. The Sc-Hf phase diagram has been taken from Ref. [61] and extended to temperatures below 1200 K by a hcp A3 single phase field.

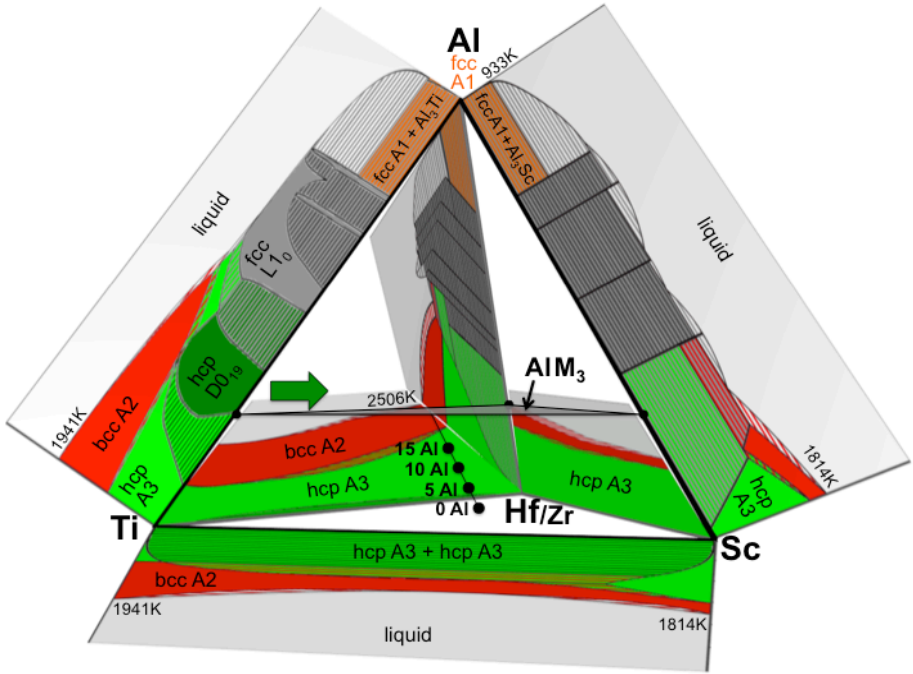

Fig. 2: 3D representation of the multi-component phase diagram constructed from the binary diagrams (Fig. 1). For visualization purposes, the Zr-binary phase diagrams (similar to the ones of Hf, see Fig. 1) are not shown. Black filled circles indicate the phase diagram trajectory for varying Al-concentration including the investigated alloys containing 5 and 15% Al.

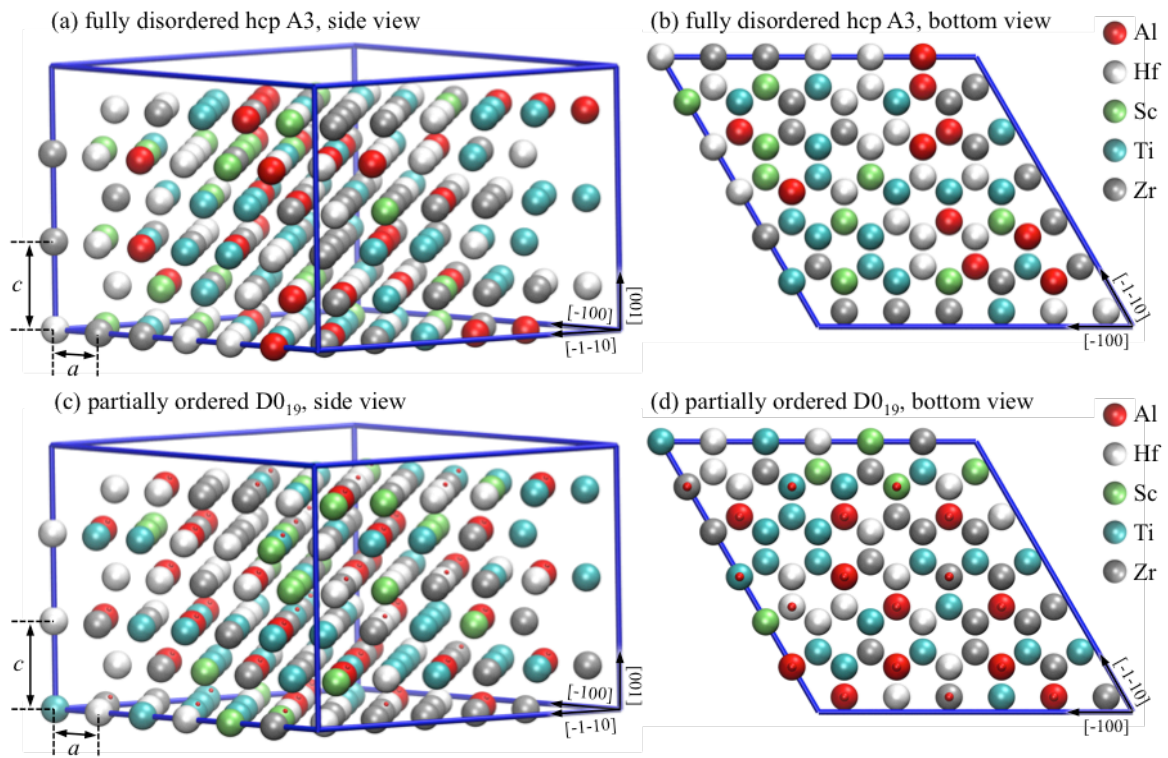

Fig. 3: Side and bottom view of the 216-atom super cells of the disordered hcp A3 structure [(a) and (b)] as well as for the partially ordered $D0_{19}$ structure [(c) and (d)] employed in the first-principles calculations for $Al_{15}Hf_{25}Sc_{10}Ti_{25}Zr_{25}$ at.%. The small red dots in (c) and (d) indicate the sublattice positions for the ordered Al atoms in the $D0_{19}$ superstructure.

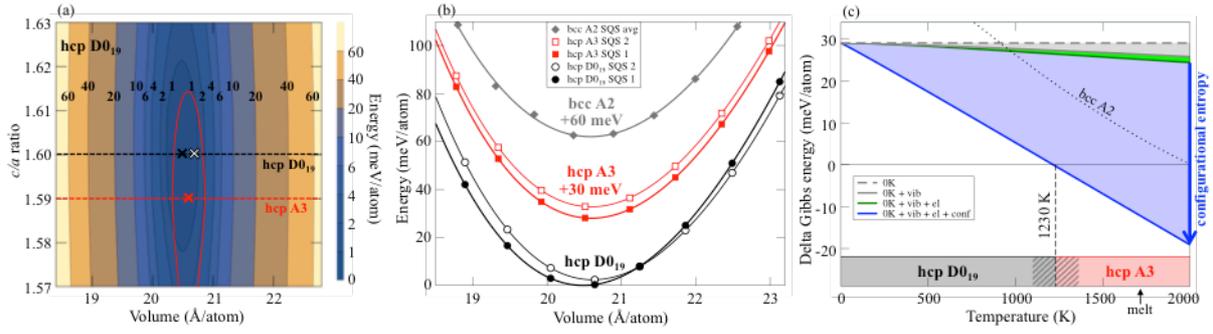

Fig. 4: *Ab initio* energetics of the hcp based phases D0$_{19}$ and A3, and of the bcc based A2 phase for the Al$_{15}$Hf$_{25}$Sc$_{10}$Ti$_{25}$Zr$_{25}$ at.% HEA. (a) Energy contour plot at $T=0$ K for the D0$_{19}$ phase as a function of volume and *c/a* ratio. The bold black numbers give the value of the contours in meV/atom. The black dashed line indicates the equilibrium *c/a* ratio (of SQS 1) and the black and white crosses mark the equilibrium volumes of the two employed supercells (SQS 1 and 2). The red solid line shows the 1 meV/atom contour of the hcp A3 energy surface. The red dashed line indicates the equilibrium *c/a* ratio for hcp A3 and the red cross the corresponding equilibrium volume. (b) $T=0$ K energy-volume curves for D0$_{19}$ and A3 (with two SQS' for each phase) at the equilibrium *c/a* ratios as indicated by the dashed lines in (a). The energy-volume curve of bcc A2 which has been obtained by an average over the two studied SQS supercells (see Sec. 2) is also shown. (c) Difference in the Gibbs energies of D0$_{19}$ and A3 as a function of temperature (D0$_{19}$ corresponding to the zero line). The Gibbs energy difference is decomposed into the $T=0$ K energy (0K), the vibrational (vib), electronic (el), and configurational (conf) contribution. The final curve corresponds to the blue solid line, resulting in a transition temperature of 1230 K as indicated by the vertical dashed line. The dotted line shows the difference between the Gibbs energy of D0$_{19}$ and bcc A2 containing all contributions. The onset of melting as obtained from our experimental results is indicated by the black arrow.

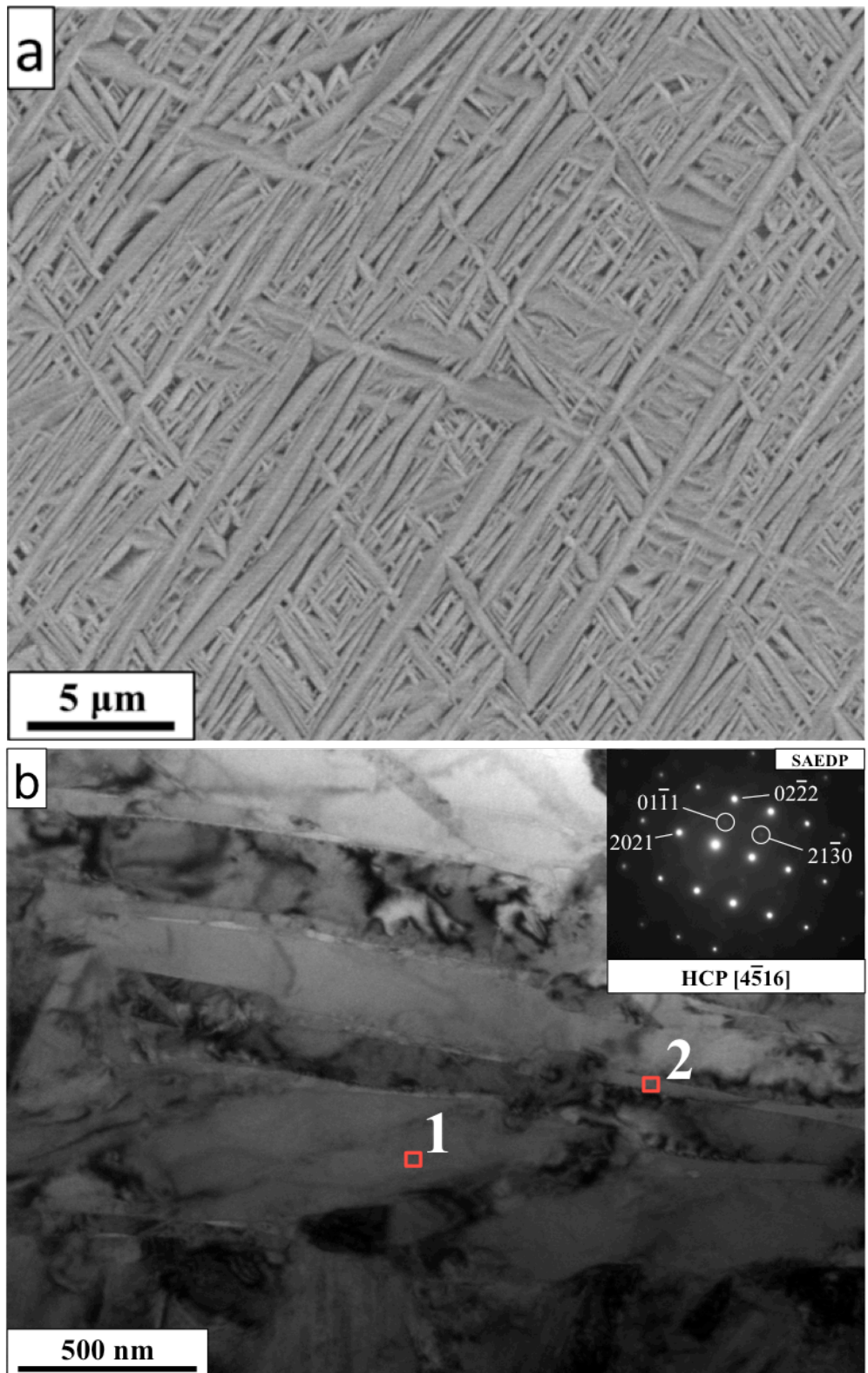

Fig. 5: Microstructure of the $Al_{15}Hf_{25}Sc_{10}Ti_{25}Zr_{25}$ at.% HEA in the as-cast state; a) SEM image, b) TEM-BF with SAED patterns and EDS point analysis taken at the red squares, with the results given in Tab. 3.

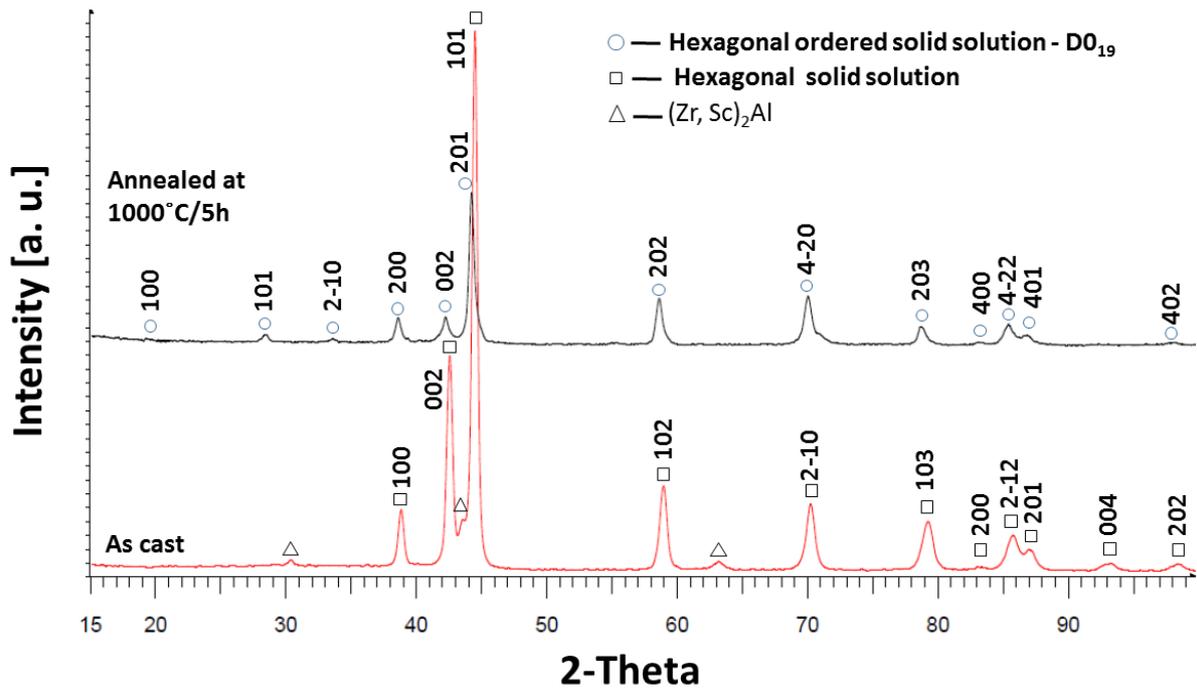

Fig. 6: X-ray analysis of the $Al_{15}Hf_{25}Sc_{10}Ti_{25}Zr_{25}$ at.% HEA in the as cast state (lower panel) and after annealing (upper panel).

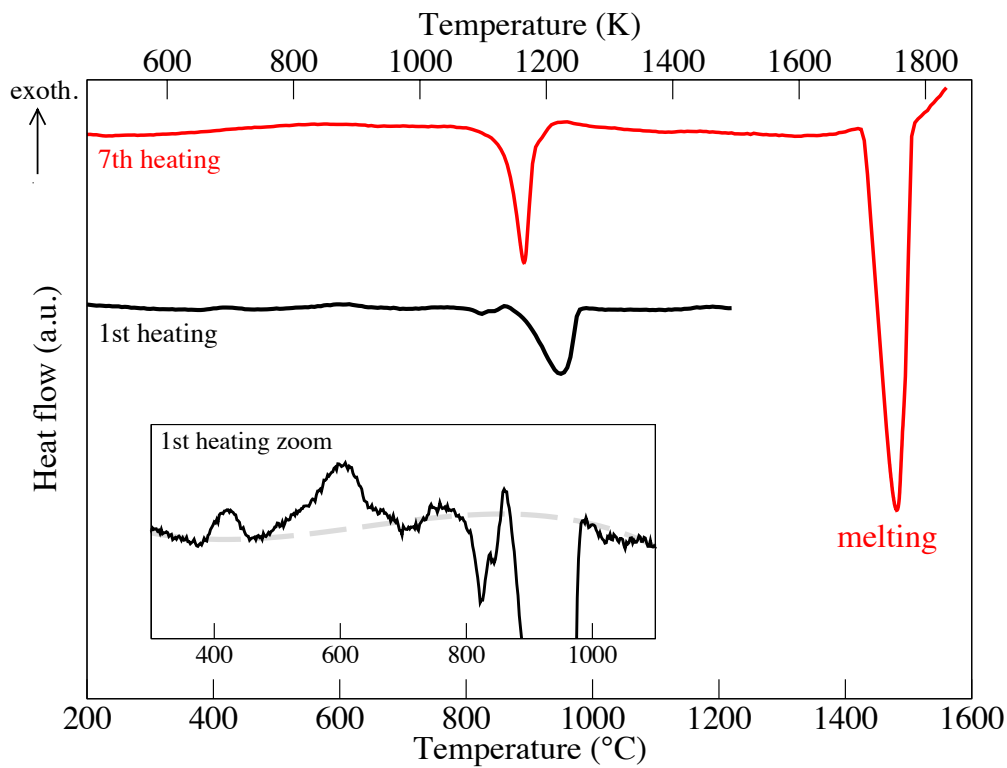

Fig. 7: DTA heat flow curves of the $Al_{15}Hf_{25}Sc_{10}Ti_{25}Zr_{25}$ at.% HEA. The black curve corresponds to the 1$^{st}$ heating cycle directly after casting. The red curve shows the 7$^{th}$ heating cycle which corresponds to the equilibrated situation, i.e., further heating cycles show the same dependence. The inset enlarges the 1$^{st}$ heating curve and the gray dashed line is a guide to the eye indicating the baseline.

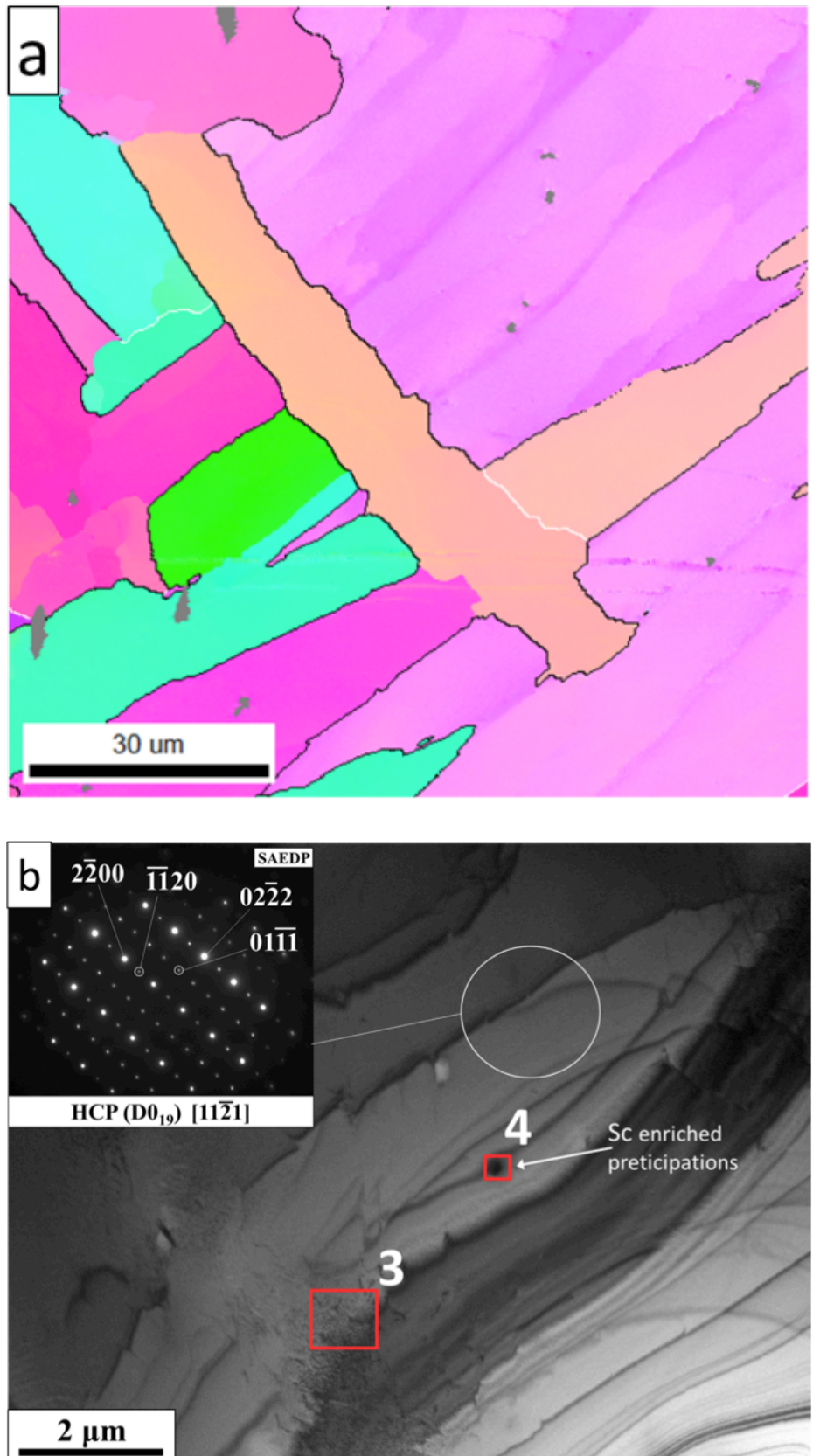

Fig. 8: a) Inverse Pole Figure (IPF) map after annealing, b) TEM bright field images with SAED patterns and points of EDS analysis after annealing (cf. Tab. 3).

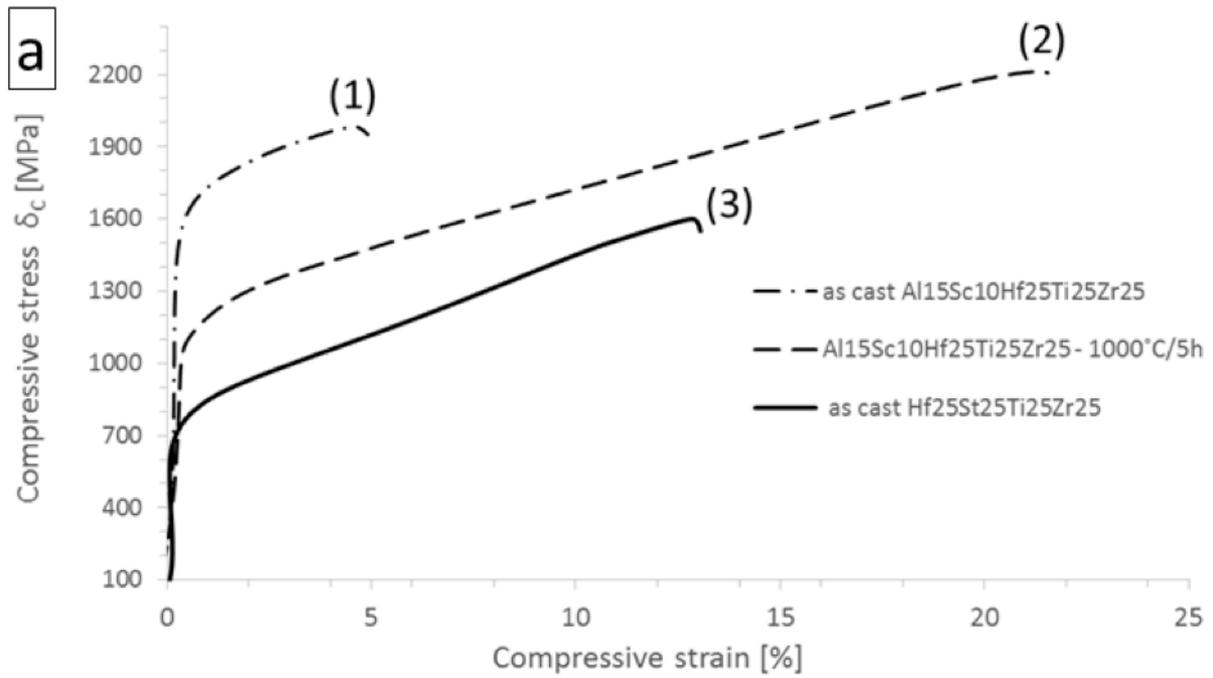

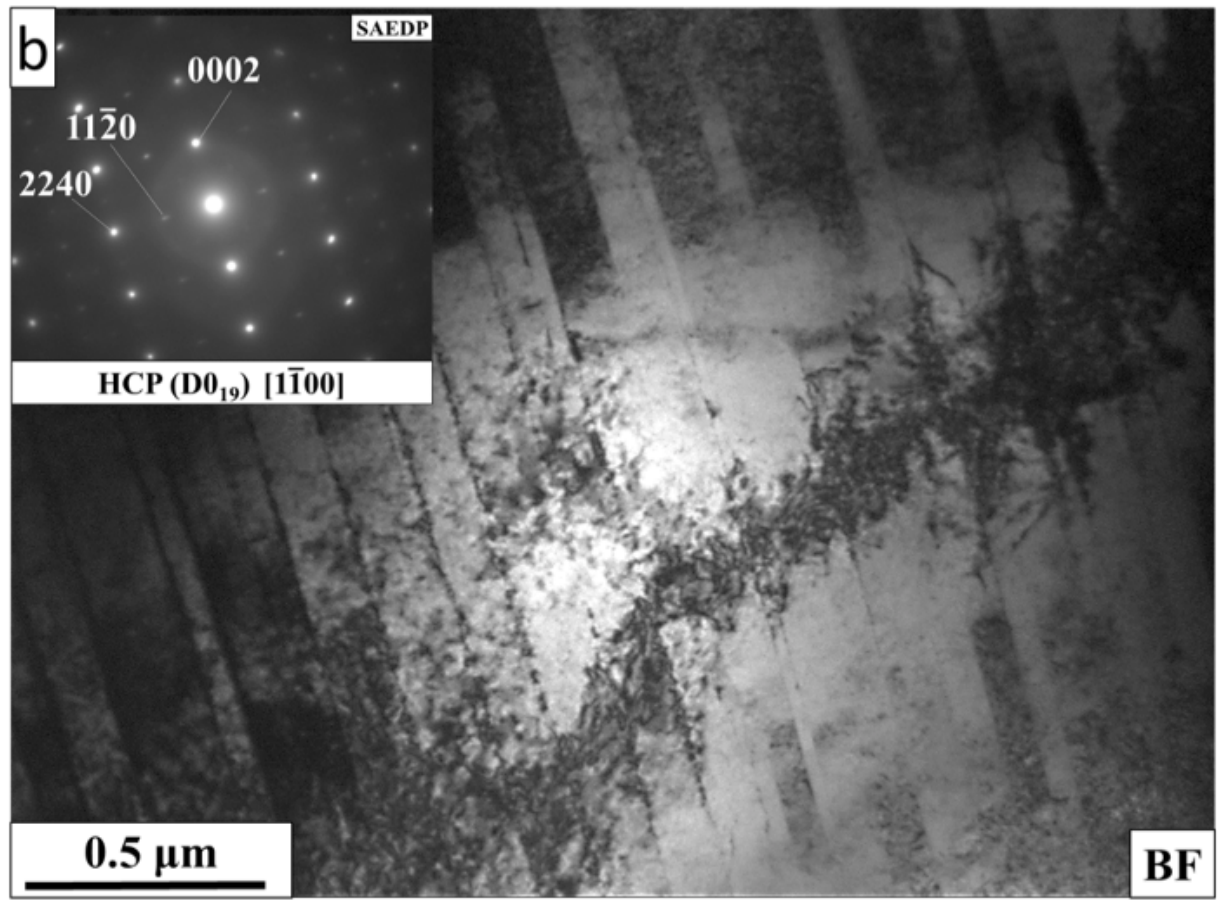

Fig. 9: (a) Compression curves of the $Al_{15}Hf_{25}Sc_{10}Ti_{25}Zr_{25}$ at.% alloy in the as cast state (curve 1) and annealed state at 1000°C for 5h (curve 2), and of $Hf_{25}Sc_{25}Ti_{25}Zr_{25}$ at.% in the as cast state (curve 3). (b) TEM bright field image with SAED pattern of the annealed $Al_{15}Hf_{25}Sc_{10}Ti_{25}Zr_{25}$ at.% HEA after straining to 22%.

|  | Al | Ti | Sc | Hf | Zr |
|---|---|---|---|---|---|
| Atomic volume (Å$^3$) | 16.6 | 17.8 | 25.0 | 22.3 | 23.3 |
| Electronegativity (Pauling scale) | 1.61 | 1.54 | 1.36 | 1.30 | 1.33 |

Tab. 1: Atomic volumes and electronegativities of the constituent elements of the investigated HEA as obtained from the CRC handbook of elements [62]. The atomic volumes have been calculated from the lattice parameters of the stable phases at ambient conditions.

|  |  | *a* (Å) | *c/a* ratio | *V* (Å$^3$) | *B* (GPa) | *ΔE* (meV) |
|---|---|---|---|---|---|---|
| hcp D0$_{19}$ | Experiment | 3.1126 | 1.593 | 20.80 |  |  |
|  | Theory (DFT) | 3.0956 | 1.60 | 20.55 | 96 | 0 |
| hcp A3 | Experiment | 3.1092 | 1.585 | 20.63 |  |  |
|  | Theory (DFT) | 3.1042 | 1.59 | 20.59 | 95 | +30 |
| bcc A2 | Theory (DFT) | 3.4527 | 1 | 20.58 | 86 | +60 |

Tab. 2: Measured and theoretically predicted values for the lattice parameter *a*, *c/a*-ratio, and volume per atom *V* for the Al$_{15}$Hf$_{25}$Sc$_{10}$Ti$_{25}$Zr$_{25}$ at.% HEA. The experiments have been performed at ambient temperature whereas the theoretical DFT values correspond to 0 K. Additionally, the theoretical 0 K values for the bulk modulus *B* and the energy difference *ΔE* with respect to the D0$_{19}$ phase are given.

|  |  | Al | Hf | Sc | Ti | Zr |
|---|---|---|---|---|---|---|
| Target |  | 15.0 | 25.0 | 10.0 | 25.0 | 25.0 |
| As cast | Coarse laths | 16.9 | 28.7 | 8.9 | 21.4 | 23.9 |
|  | Interspace region | 20.6 | 21.9 | 8.5 | 24.1 | 24.8 |
| Annealed | Laths | 16.1 | 27.0 | 8.3 | 21.1 | 26.6 |
|  | Precipitates | 1.6 | 0.0 | 97.8 | 0.2 | 0.3 |

Tab. 3: TEM-EDS measured concentrations (at.%) of the as cast alloy in the two regions indicated in Fig. 5b by the red squares, corresponding to a region inside the coarse laths (label 1) and to the interspace region between the (coarse and fine) laths (label 2) as well as for the annealed sample shown in Fig. 7a corresponding to a lath (label 3) and precipitate (label 4).

# Computational engineering of sublattice ordering in a hexagonal AlHfScTiZr high entropy alloy

Lukasz Rogal[1a], Piotr Bobrowski[1], Fritz Körmann[2], Matthias Wegner[4], Daniel Gaertner[4], Sergiy Divinski[4], Gerhard Wilde[4], Frank Stein[3], Blazej Grabowski[3b]

[1]*Institute of Metallurgy and Materials Science of the Polish Academy of Sciences, 30-059 Krakow, Poland*
[2]*Materials Science and Engineering, Delft University of Technology, 2628 CD Delft, Netherlands*
[3]*Max-Planck-Institut für Eisenforschung GmbH D-40237 Düsseldorf, Germany*
[4]*Institute of Materials Physics, University of Münster, Wilhelm-Klemm-Str. 10, 48149 Münster, Germany*

*Corresponding authors:* [a]*L. Rogal, l.rogal@imim.pl, tel.+48 122952801, fax +48 122952804*
[b] *Blazej Grabowski, b.grabowski@mpie.de, tel:+49 211 6792 512, fax:+492116792512*

**Micro-mechanical tests for $Al_{15}Hf_{25}Sc_{10}Ti_{25}Zr_{25}$ at.%**

Tensile tests were performed on a custom-built device. Dogbone-shaped samples with a gauge length of 3 mm and 1.0x0.3 mm$^2$ were cut by spark erosion from as-cast samples and from a sample after thermal treatment at 1000°C for 5h.

The results of the tensile tests are shown in Fig. S1. In the as-cast state (red line) the yield strength approaches a value of about 1400 MPa and the elongation to failure (excluding the elastic part) is about 0.5%, Fig. S1. As for the annealed state, two samples were investigated (green and blue curves) to improve statistics. Both samples revealed brittle fracture at a stress of about 1000 MPa.

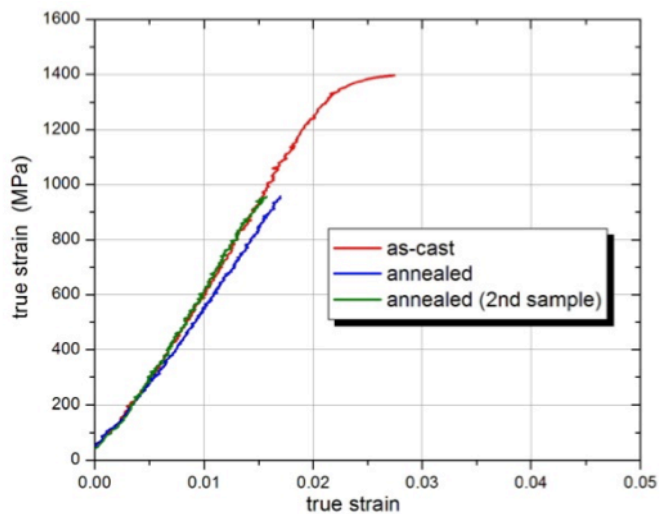

**Figure S1**. True stress-true strain tensile curves for the as-cast (red line) and annealed (green and blue lines) $Al_{15}Hf_{25}Sc_{10}Ti_{25}Zr_{25}$ at.% HEA.

Although our custom-built device does not provide correct values of the elastic modulus, the yield strength and elongation to failure are reliable [1].

The fracture surfaces were examined by a FEI NanoSEM 230 field-emission-gun scanning electron microscope (FEG-SEM). The results are presented in Fig. S2. The as-cast sample reveals a pronounced ductile fracture with a combination of transgranular and intergranular features and a developed vain-like structure, Fig. S2a and b. In contrast, the annealed samples reveal brittle fracture with predominantly transgranular character, Fig. 2c, d. Nevertheless Fig. 2c and d document a pronounced dislocation activity during fracture of the ordered Al$_{15}$Hf$_{25}$Sc$_{10}$Ti$_{25}$Zr$_{25}$ at.% HEA.

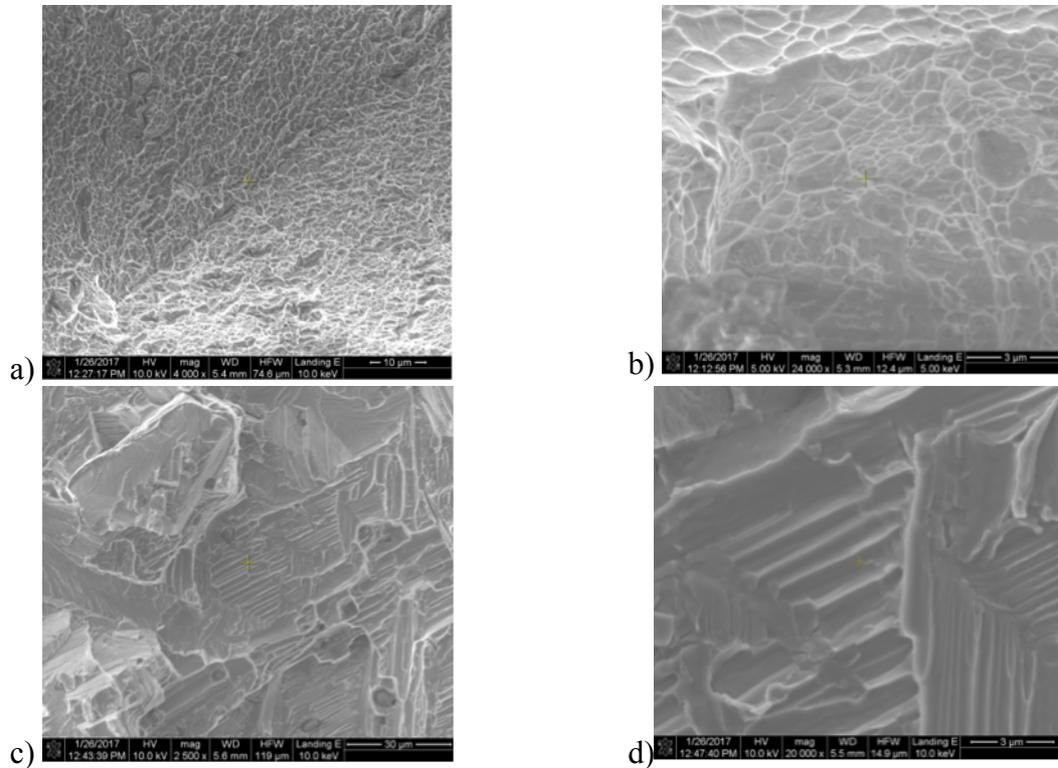

**Figure S2**: SEM images of the fracture surfaces observed on the as-cast (a, b) and annealed (c, d) Al$_{15}$Hf$_{25}$Sc$_{10}$Ti$_{25}$Zr$_{25}$ at.% HEAs.